\documentclass[doublecolumn,journal]{IEEEtran}

\usepackage{epsfig}
\usepackage{amsmath,mathrsfs,txfonts}
\usepackage{amssymb}
\usepackage{graphicx}
\usepackage{multirow}
\usepackage{morefloats}
\usepackage{subfigure}
\usepackage{color}
\usepackage{caption}

\begin{document}
\title{Signal Identification for Adaptive Spectrum Hyperspace Access in Wireless Communications Systems}

\author{\IEEEauthorblockN{Ali Gorcin\IEEEauthorrefmark{1},
Huseyin Arslan\IEEEauthorrefmark{1}\IEEEauthorrefmark{2},\\
\IEEEauthorblockA{\IEEEauthorrefmark{1}Department of Electrical Engineering,
University of South Florida, 4202 E. Fowler Ave., ENB-118, Tampa, FL, 33620, USA}\\
\IEEEauthorblockA{\IEEEauthorrefmark{2}Faculty of Engineering and Architecture, Istanbul Medipol University, Kavacik Mah. Ekinciler Cad. No.19 Kavacik Kavsagi,
Beykoz, Istanbul, 34810, Turkey}\\
Emails: agorcin@mail.usf.edu, arslan@usf.edu}}

\maketitle

\begin{abstract}
Technologies which will lead to adaptive, intelligent, and aware wireless communications systems are expected to offer solutions to the capacity, interference, and reliability problems of wireless networks. The spectrum sensing feature of cognitive radio (CR) systems is a step forward to better recognize the problems and to achieve efficient spectrum allocation. On the other hand, even though spectrum sensing can constitute a solid base to accomplish the reconfigurability and awareness goals of next generation networks, a new perspective is required to benefit from the whole dimensions of the available electro (or spectrum) hyperspace, beyond frequency and time. Therefore, spectrum sensing should evolve to a more general and comprehensive awareness providing mechanism, not only as part of CR systems but also as a communication environment awareness component of an adaptive spectrum hyperspace access (ASHA) paradigm which can adapt sensing parameters autonomously to ensure robust signal identification, parameter estimation, and interference avoidance. Such an approach will lead to recognition of communication opportunities in different dimensions of the spectrum hyperspace, and provide necessary information about the air interfaces, access techniques and waveforms that are deployed over the monitored spectrum to accomplish ASHA, resource and interference management.
\end{abstract}
\IEEEpeerreviewmaketitle
\section{Introduction}
\label{INT}
Wireless communications systems undergo an evolution as voice oriented applications evolve to data and multimedia based services. Eventually, the progress of wireless technologies and standards from the first generation through the fourth and beyond lead the prevalence of wireless services among daily users by triggering an extensive growth on the demand for wireless communications services. Along with the rapid rise at the number of users, the increasing demand for more communications capacity to deploy multimedia applications entail effective utilization of communications resources. As a matter of fact, this requirement stems from the limited resources such as frequency spectrum, physical limits on communications, data transmission limitations indicated by Shannon - Hartley Theorem, and the characteristics of the wireless channel. When the current state of wireless technologies is considered, improvements should be achieved in terms of 
\begin{itemize}
\item effective spectrum allocation,

\item adaptive and complex modulation, error recovery, channel estimation, diversity and code design techniques to allow high data rates while maintaining desired quality of service (QoS),

\item reconfigurable and flexible air interface technologies for better interference and fading management, and 

\item cooperation of these concepts in an environment that they exist along with the present wireless technologies. 
\end{itemize}

Traditional communications systems are designed to allocate fixed amounts of resources to the users. These systems do not employ adaptive resource utilization techniques. For instance the federal communications commission (FCC) frequency allocation chart indicates scarcity of frequency bands, especially at the ultra-high frequency (UHF) range for the United States of America. Based on the information acquired from the chart and the traditional communications approach, a new spectrum allocation auction at the upper UHF bands can be seen as a straightforward answer to the problem. However more auctions would be a temporary solution, because the scarcity problems will re-emerge by the time, as the number of users grow. Secondly the wireless propagation characteristics complicate the implementation of communications systems at the super and extremely high frequency bands and coverage problems emerge. Driven by these issues, researchers inclined to evaluate the efficiency of the wireless systems deployments in the spectrum. Extensive measurement and data analysis activities in the last decade indicated a significant difference between real spectral utilization and chart allocations because, while the static transmitters such as digital tv signals continuously occupy the spectrum, the dynamic users such as land mobile radio systems transmit intermittently and do not occupy the spectrum continuously.  Therefore, it is a spectrum allocation inefficiency problem rather than a scarcity issue but it was not possible to observe this solely based on spectrum allocation charts and traditional communications approaches. As a result, the dynamic usage of the spectrum must be distinguished from the static usage, because the intermittent utilization of the spectrum by the dynamic users imply new \textit{communications opportunities} to be exploited to access the spectrum. Adaptive wireless communications design methodologies are proposed to identify the requirements of the users and to allocate just enough resources consistently. These methodologies enable more efficient utilization of the system resources and consequently improve the total capacity usage. The dynamic spectrum access (DSA) paradigm is one of these adaptive communications methodologies which is proposed to achieve efficient frequency spectrum utilization.

Among the definitions of radio systems which are computationally intelligent \cite{Fettweis96}, widely accepted terminology to accomplish efficient frequency spectrum utilization is introduced through software defined radios (SDR) and later on cognitive radios \cite{Mitola99}. Cognitive radio (CR) technology aims to choose and support multiple variations of wireless communications systems, and introduces secondary (or sometimes unlicensed) users to achieve opportunistic access in the wireless spectrum. Allocation of secondary users should be conducted in such a reliable and flexible way that the communication of the primary (licensed) users would not be affected and there should be no loss at the QoS due to the secondary access. To this end, spectrum sensing is proposed as a CR feature to detect the primary users via digital signal processing methods such as energy detection, matched filtering, covariance matrix based algorithms, cyclostationary feature detection, and multi-taper spectral estimation \cite{Axell}.

Initial spectrum sensing methods are mostly designed assuming that only a single channel is sensed. However, if a block or multiple channels of the wireless spectrum are sensed instead of a single channel, more communications opportunities will become available. Therefore, wideband sensing approach is conceived as an expansion of spectrum sensing over the wireless spectrum and CR should focus on a wideband spectrum, if possible. On the other hand, sensing methods mostly assume that the channel frequency response is flat, but this assumption does not hold for wideband and multi-channel scenarios. In addition, the binary hypothesis testing which only decides whether there is an occupant signal in a given channel or not is frequently utilized by sensing methods. When binary hypothesis testing is directly applied to the wideband, multi-channel scenarios, decision mechanism can lead to the assumption of full occupancy even though the spectrum is not fully occupied since the CR will not be able to discriminate multiple channels from a single channel. Eventually, spectrum sensing methods cannot be directly applied to the wideband scenarios and consequently, wideband sensing techniques such as multiband joint detection, wavelet, sweep tune, filter bank detection are proposed in the literature \cite{Hongjian}.

When the wideband sensing techniques are considered, multiband joint detection and wavelet detection techniques require high sampling rate/high resolution analog to digital conversion (ADC) blocks which are computationally expensive and difficult to implement. However, the rest of the sensing procedure is achieved simply either by dividing the time data into the bins and applying energy detection over the power spectral density (PSD) of each bin, or by applying wavelet transform over the PSD. On the other hand, the sweep tune approach benefits from the superheterodyne receiver architecture and sweeps over the frequencies of interest by mixing the output of a tunable local oscillator with the input signal and down-converting it to intermediate frequency range. After a bandpass filter is applied to the channel of interest, spectrum sensing methods are utilized. This method is slow and its sequential monitoring and sensing approach can lead to loss of significant spectrum occupancy information. In filter bank detection, a set of filters are designed to process the wideband signal and each filter down converts the corresponding block of the spectrum. Then, sensing methods can be applied over each block separately. The main drawback of this technique is the complexity of the parallel filter bank architecture implementation. Finally it should be mentioned that the signal processing problems caused by the difficulty of providing a certain dynamic range \textit{i.e.}, high resolution sampling, while maintaining the Nyquist sampling rate led to a new set of sensing techniques. The ADC is conducted in sub-Nyquist rates and then spectral reconstruction is applied on the sampled signal. Sub-Nyquist techniques provide a solution to the high speed ADC, however, the monitored spectrum should be sparsely occupied for successful sensing. Maintaining this criterion can be difficult in many scenarios therefore some improvements at the sensing performance should be expected in the sub-Nyquist domain.

\section{Motivation}
\label{MOT}
\begin{figure*}[hbt]
\centering
\includegraphics[width=\textwidth]{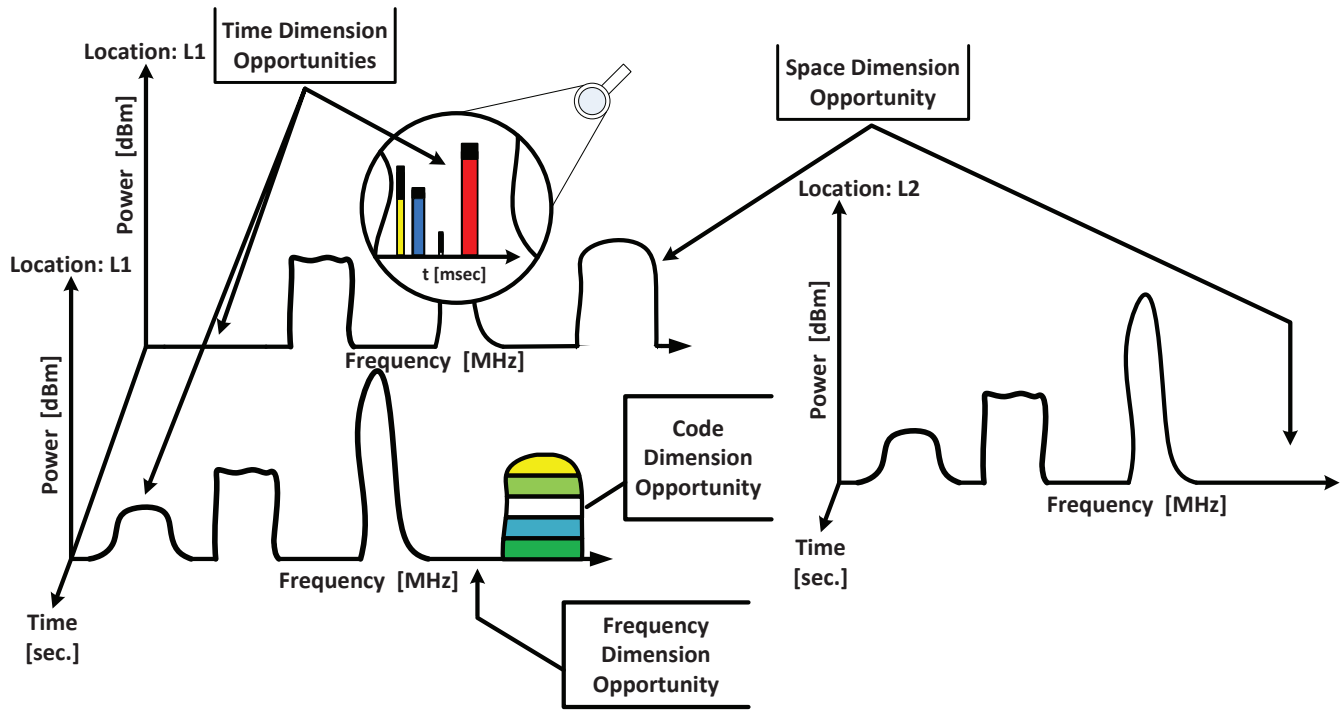}
\caption{Spectrum hyperspace provides communications opportunities in space, time, code, and other dimensions.}
\label{c1fig2}
\end{figure*}

Spectrum sensing is perceived as a feature that solely provides whether the primary user exists in a communication channel or not. Its initial scope can be defined as an information provider in time or frequency domain, on the channel access status\footnote{Channel term here refers to a frequency band of a certain bandwidth as defined in \cite{Qing11}.} of the primary users to enable secondary access \cite{Qing11}. Extension of sensing to the wideband scenarios and the final decision stages of the current wideband sensing techniques which are based on binary hypothesis testing also prove this perception \cite{Hongjian}.  However, the FCC Spectrum Policy Task Force report \cite{FCC01} recommends the improvement of wireless throughput by achieving signal orthogonality over a range of dimensions grouped under a domain called as \textit{spectrum hyperspace} \cite{Horne}. Spectrum hyperspace includes but not limited to frequency, time, space, power, polarization, angle of arrival, and code dimensions. Secondary access should be extended to these dimensions. In Fig.~\ref{c1fig2} some of the communications opportunities provided by the spectrum hyperspace are illustrated.

Utilization of the dimensions of the hyperspace, to improve the throughput, not only requires the spectrum sensing methods to decide whether the primary users exist in the channel or not but also necessitates identification of the primary user waveforms \textit{e.g.}, the air interface parameters, burst structures, chip rates, cyclic prefix sizes, preambles and employed radio access techniques (RATs) without going to demodulation stage. In fact, the International Telecommunication Union Radiocommunication Sector (ITU-R) report on definitions of SDR and CR \cite{SM.2152} indicates that CR should ``dynamically and autonomously adjust its operational parameters and protocols according to its obtained knowledge in order to achieve predefined objectives''. Moreover, ITU-R reports on CR systems in the land mobile services and international mobile telecommunications (IMT) systems indicate the usage of CR technology to improve the management of assigned spectrum resources. The ITU-R reports describe CR as the enabler of opportunistic spectrum access amongst wireless network operators without any prior agreements, and propose CR as the controller of the terminal reconfiguration in heterogeneous networks \cite{M.M.2225,M.M.2242}. One example area that will benefit from these definitions is the channel usage prediction methods such as time domain opportunistic channel allocation techniques because they accomplish DSA by exploiting the idle periods between bursty transmissions based on the signal identification information \cite{Geirhofer}. Another case is the maximum likelihood (ML) signal direction of arrival (DOA) estimators. Instead of being unknown stochastic processes, if the incident signals are known, useful DOA estimation properties can be derived peculiar to ML estimators \cite{Rappaport}. Therefore, CR should be aware of the communications environment as much as possible to achieve the rest of the goals itemized above. To this end, sensing procedures should provide the information about the wireless signals in the communication medium for CR to adjust the transmission parameters adaptively \cite{Gueguen22,Bjorsell}.

When the research on the adaptive and dynamic access to the wireless spectrum is considered, it is seen that the reconfigurability and awareness issues are initially addressed in the context of the transparently reconfigurable ubiquitous terminal (TRUST) project, and a radio mode identification and switching concept is introduced. Mode identification can be conducted blindly only by the radio or in an assisted manner when a priori information is available \cite{Farnham333}. The mode identification concept defined in the TRUST project later on extended to initial mode identification and alternative mode monitoring methods in \cite{Mehta777} and identification of RATs are accomplished based on received signal strength indicators. On the other hand, a configurable receiver architecture that classifies the wireless signals based on bandwidth estimation with the radial basis function neural networks is introduced in \cite{Palicot888}. In addition, classification of overlapping air interfaces using pattern recognition techniques over distributed terminals is proposed in \cite{Gandetto999}. While an air interface identification system that employs cyclostationary feature detection is defined in \cite{Menguc1000}, pattern classification and machine learning methods are combined to classify wireless signals based on characteristics such as burst size, hopping pattern, and carrier number in \cite{Mody111}. The proposed methodology introduces learning and prediction functionalities therefore it is possible to classify new signals and extend the communications capability of the proposed CR based system. Spectrum power measurements with low temporal resolution are utilized to achieve machine learning supervised classification of the RATs in \cite{Gueguen22}. A communications module aiming to attain a certain description of the radio environment by the estimation of signals modulation type, symbol rate, carrier frequency, and pulse shaping is described in \cite{Bjorsell} with the technical focus on modulation identification. Moreover, a detailed analysis of cyclostationary feature detection based modulation identification is given in \cite{Dobre}. Finally, two-stage sensing mechanisms which apply two narrow-band sensing methods sequentially are proposed to improve the overall sensing performance \cite{Hur,Yue}. 

Whether under the concept of CR or not, the research on extracting more information from the communications medium focuses on some different and specific aspects of adaptiveness and awareness requirements of next generation communications networks, instead of providing a comprehensive and unifying approach. Moreover, even though spectrum sensing can constitute a solid base to achieve the reconfigurability and awareness goals described, current technical level of sensing cannot satisfy the requirements of fully adaptive, aware and intelligent communications systems. For instance, instead of providing the extensive set of signal parameters listed above, current wideband sensing methods are designed to operate as parallel or sequential multi-channel energy detectors. In its current perception, wideband sensing can only inform CR whether the sensed channels are occupied or not based on some a priori information such as noise variance of each sensed channel. Therefore, spectrum sensing should evolve to a more general and comprehensive awareness providing mechanism, which will not only be a part of CR systems to provide channel occupancy information but also will become a communication environment awareness component of an adaptive spectrum hyperspace access (ASHA) paradigm, beyond DSA. ASHA should

\begin{itemize}
\item achieve initial autonomous signal identification and adapt sensing parameters continuously to ensure robust identification of the signals,

\item decide whether the employment of further techniques such as statistical channel occupancy prediction, mobility level estimation, user localization is necessary before accessing the hyperspace and in case of positive decision, execute these tasks,

\item select the optimized waveforms and access techniques based on the collected information on the communication medium, taking different scenarios such as secondary access, low or high priority primary access, interference prohibited or restricted access into account,

\item access the spectrum hyperspace adaptively, monitor the activities in the hyperspace to ensure the access is achieved in the predefined context and constantly evolve resource management and optimization,
\end{itemize}

\noindent and in such a perspective, spectrum sensing evolves to \textit{signal identification} which leads to recognition of communications opportunities in different dimensions of the hyperspace, and provides necessary information about the air interfaces, access techniques and waveforms to accomplish ASHA in 5G networks, CR systems, small cell and heterogeneous network terminals. 

\section{Signal Identification for ASHA}
\label{SIDSA}  
\begin{figure*}[!hbt]
\centering
\includegraphics[width=\textwidth]{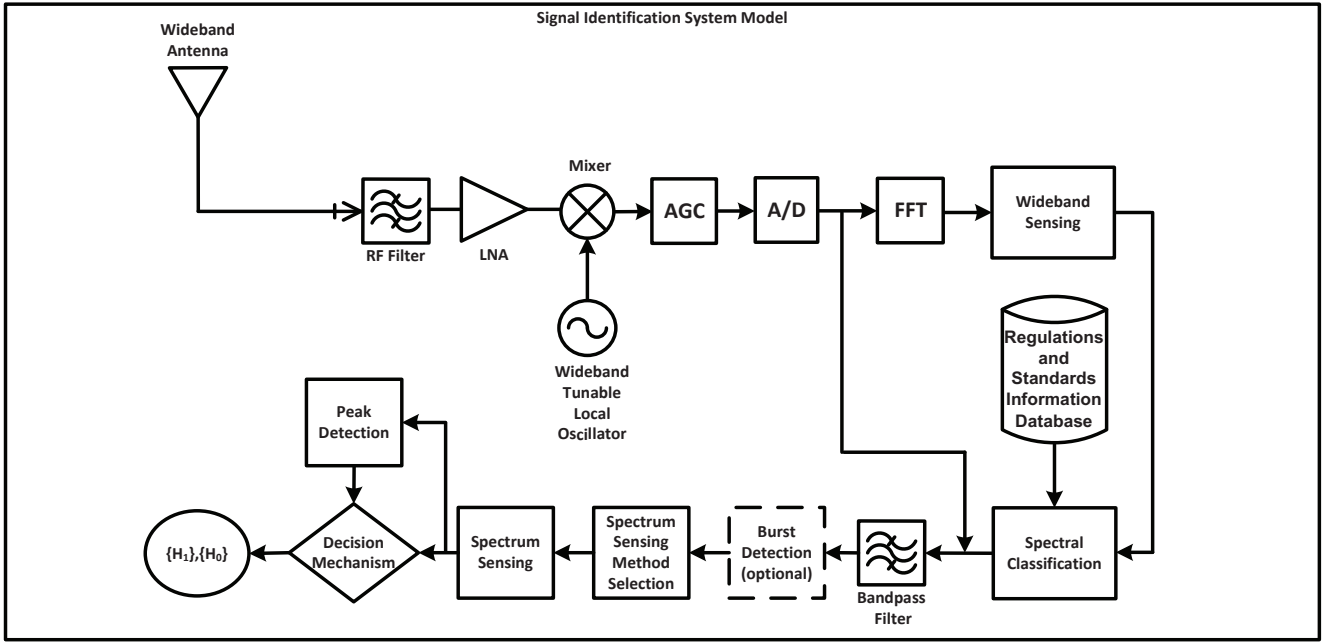}
\caption{Signal identification system model. Initial wideband sensing stage aims spectral detection and parameter estimation. Final identification is achieved based on unique features of wireless signals.}
\label{c1fig3}
\end{figure*}

We define signal identification as a procedure, which not only provides the information whether the spectrum hyperspace dimensions of interest are occupied or not, but also reveals the underlaying information regarding the parameters, such as employed channel access methods, duplexing techniques and other parameters related to the air interfaces of the signals to satisfy the ASHA requirements. To achieve signal identification, a comprehensive signal detection, sensing and classification approach is required and in this paper, we propose the signal identification system model given in Fig.~\ref{c1fig3}. In this system model, after the wideband receiver antenna, an initial RF filter which blocks high-frequency signals from the mixer input and prevents mixer non-linearities is utilized. Following the RF front-end, the features of the ADC possess crucial importance for the performance of the signal identification procedure. Signals that have high power levels are mixed with very low power signals in the spectrum. Therefore, digitization process should have the capability to represent a wide dynamic range in the digital domain. This requirement can be satisfied by the employment of high resolution ADC circuity such as $12$ and $16$ bit converters proposed in the literature \cite{Cabric987}, but the wideband nature of the conversion also demands high sampling rates \cite{Hongjian}. To overcome the difficulty of balancing the resolution and speed requirements, implementation of  notch filters at the RF front-end to isolate high-power signals and sampling sub-Nyquist rates can be considered, however, both approaches assume that certain information such as frequency channels of high power signals and spectral occupation rate are available beforehand. In our work, we assume that a wide dynamic range is available through a low noise amplifier (LNA) at the mixer input to achieve high level input power sensitivity. On the other hand an automatic gain control (AGC) block is allocated before the ADC to enable to the resolution adjustment of the conversion. Therefore, balance between resolution and speed can be accomplished in an adaptive manner. 

\subsection{Selection of Initial Dimension of Operation and Wideband Sensing}
\label{SIOD} 
The received signal is composed of noise and various signals which employ different access techniques and comprise unique characteristics inherited from the definition of their technologies \textit{e.g.}, standards, air interfaces. These features have projections over the multiple dimensions of the spectrum hyperspace and can be used to identify these signals. However, a certain level of dimensional abstraction is required beforehand to initialize the identification process. When the dimensions of the spectrum hyperspace are taken into account, both channel assignments over the wireless spectrum which are managed by the regulatory organizations and standard based carrier spacings imply frequency as the initial dimension to start the separation of the signals. Therefore, frequency domain representation of the wideband signal is obtained through fast Fourier transform (FFT) at the next stage of the proposed signal identification system model in Fig.~\ref{c1fig3}. It should be noted that, FFT is also the initial block of wideband sensing algorithms such as multiband joint detection and wavelet detection. 

Wideband sensing stage coarsely detects the active channels over the given spectrum and provides the bandwidth and center frequency estimation information. For instance, while multiple users can access a single channel with burst transmission, there can be a single user in a given channel using direct-sequence spread spectrum (DSSS) modulation or a single user can access multiple channels. Moreover, wireless signals can overlap in multiple dimensions leading to interference. Therefore, before starting any identification procedure, a certain level of signal separation should be achieved. Bandpass filter and the following optional burst detection blocks aim to conduct this separation operation based on the input information provided by the wideband sensing stage. If the signals are overlapped in time, bandpass filter that operate over the estimated band of interest leads to elimination of the time domain components of other channels. If there are still multiple users or techniques accessing the channel, before the selection and the execution of the spectrum sensing method, burst detection can be applied to the filtered data. On the other hand, if the sensing procedure is affected from the spectral overlap, filtering and the burst detection procedures can also provide spectral distinction.

\begin{figure*}
        \centering
        \subfigure[Recorded time domain ISM band signal.]{\includegraphics[width=0.32\textwidth]{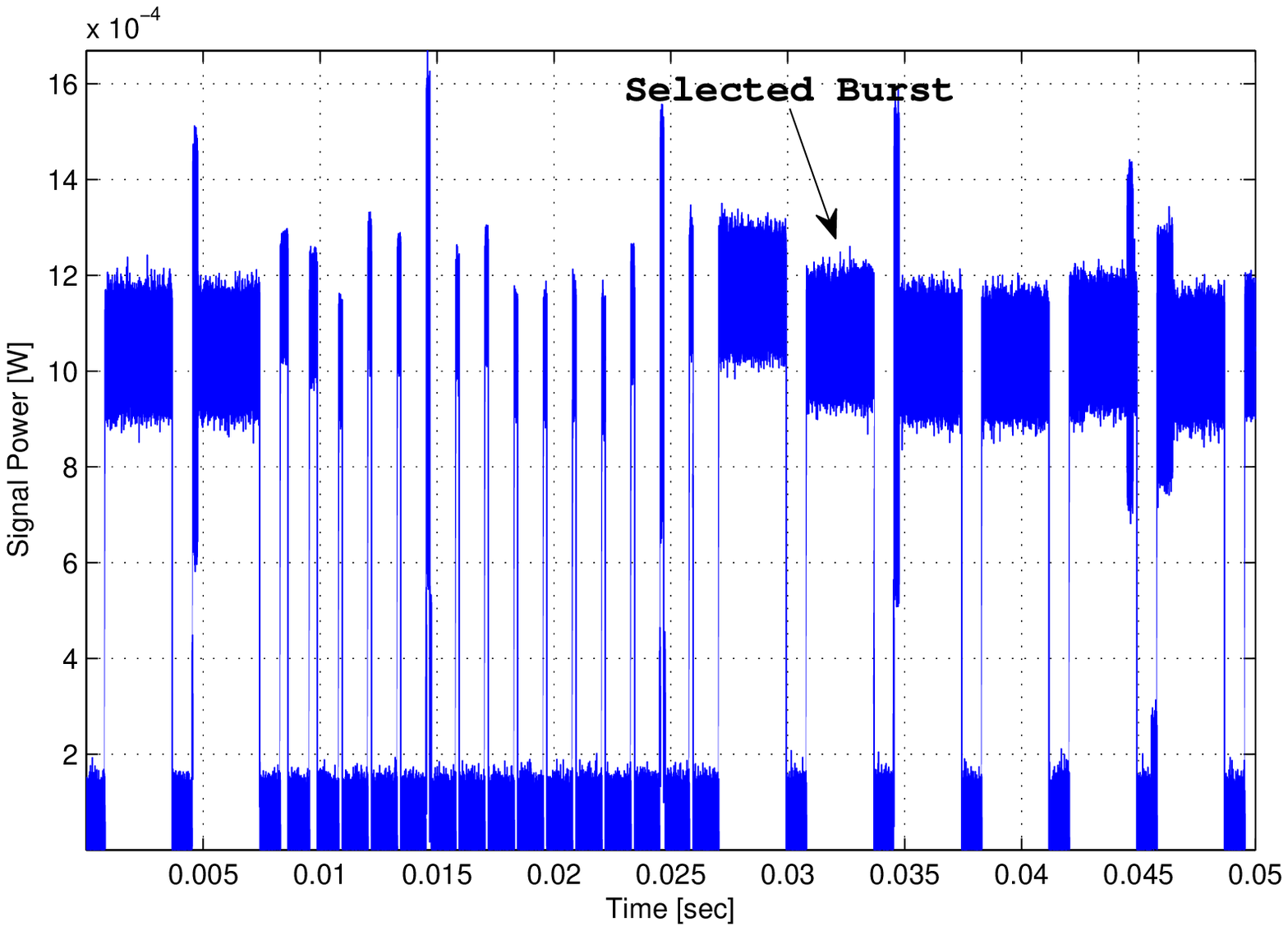}
                        \label{fig:444}}
       \subfigure[Cyclostationary feature detection without filtering.]
       {\includegraphics[width=0.32\textwidth]{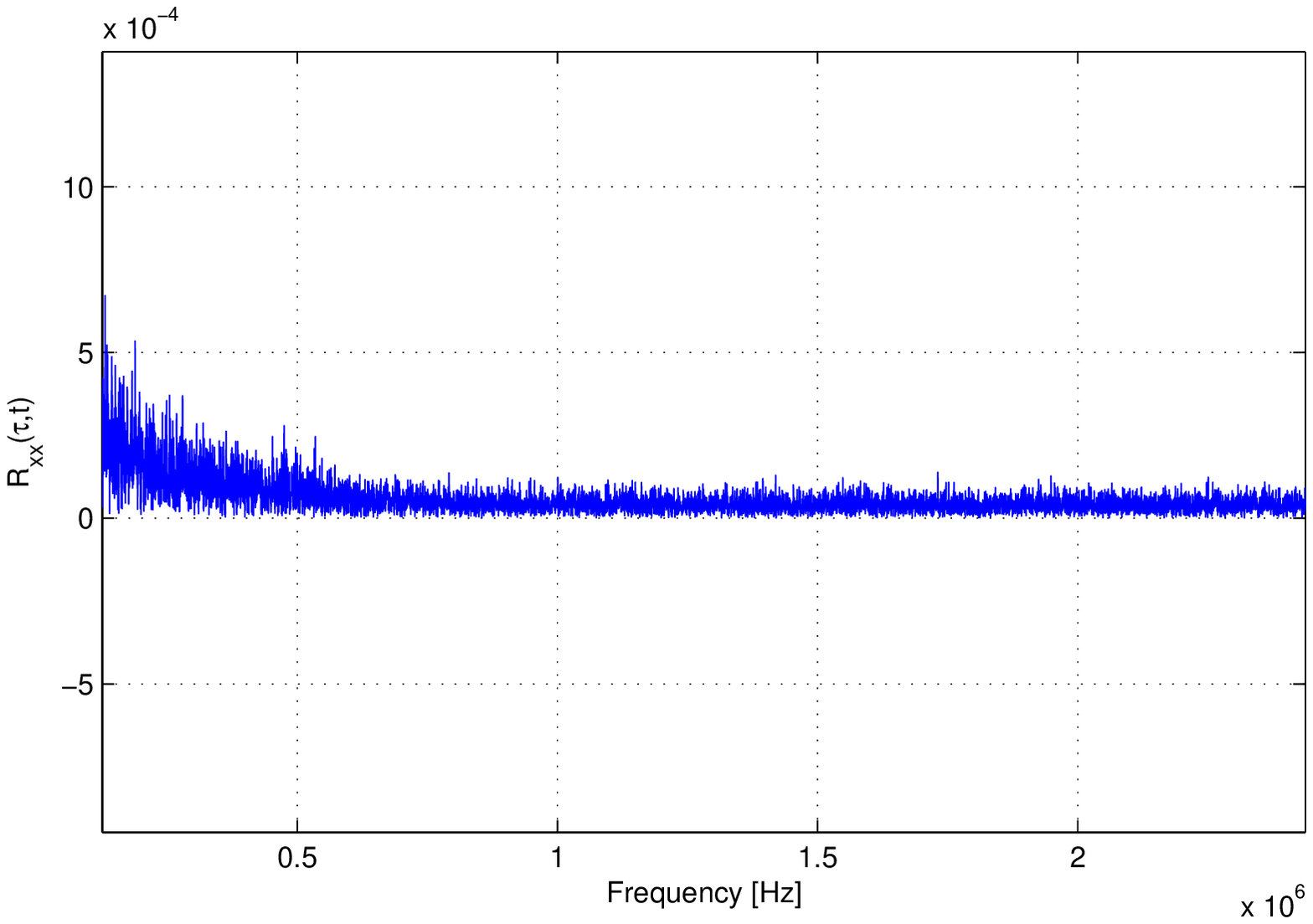} \label{fig:777}}
        \subfigure[Cyclostationary feature detection after filtering. The cyclic frequency at $1$ MHz due to burst header is visible.] {\includegraphics[width=0.32\textwidth]{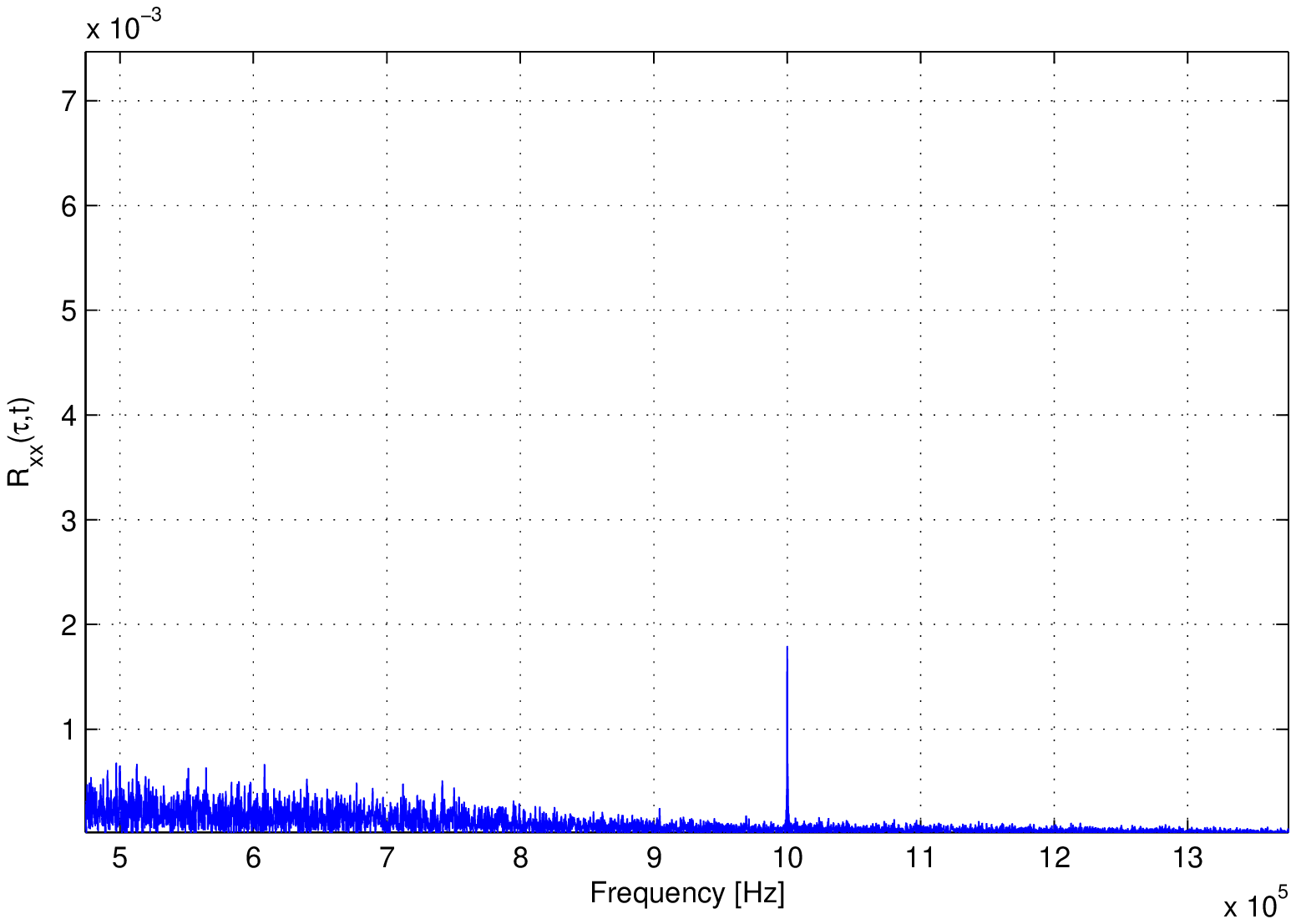}
                                \label{fig:888}}
        \caption{Effect of filtering on spectrum sensing methods. $R_{xx}(\tau, t)$ at the vertical axises represents the time varying cyclic autocorrelation function which is periodic in time $\tau = 1$ $\mu$secs for Bluetooth signals.}\label{fig:filtering}
        \label{c1fig1}
\end{figure*}

One alternative approach could be starting the identification process in time domain and designing the blocks following the ADC under this assumption. To investigate such a scenario, we conducted Bluetooth based communications beside the ongoing communications activity in the industrial, scientific and medical (ISM) band and recorded the whole 80 MHz band from  $2.4$ GHz to $2.48$ GHz with the wideband receiver available in our laboratory. Fig.~\ref{fig:444} shows the time domain data of 50 milliseconds of recording. $54$ bits fixed header of the Bluetooth signals makes them suitable for data rate estimation via cyclostationary feature detection and we verified through demodulation that the burst we marked in Fig.~\ref{fig:444} is a Bluetooth signal centered around $2.44$ GHz. But when the cyclostationary feature detection is applied, the resulting cyclic spectrum of the selected burst does not reveal the data rate information that is expected as shown at the cyclic spectrum in Fig.~\ref{fig:777}. Therefore, in a wideband sensing scenario, if the burst detection is conducted without filtering each signal, some other dominant frequencies over the detected burst may overshadow the features of the signal that would normally lead to identification. On the other hand, when the procedures of introduced signal identification system model are followed, as the bandpass filter is applied to leave only the time domain components unique to the estimated frequencies, for instance in the given peculiar scenario, final sensing process based on cyclostationary feature detection will lead to the dominant peak indicating data rate of Bluetooth signals as given in Fig.~\ref{fig:888}.

\subsection{Spectral Classification and Other Blocks}
\label{SCOB} 
Spectral classification block (SCB) is the information comparison block between the signal processing blocks of wideband sensing and bandpass filtering. After the wideband sensing stage detects the signals in the frequency domain and estimates the bandwidths and center frequencies of the detected signals, SCB compares the estimated spectral information with the regulatory and wireless standard based information and decides which candidate or potential signals are possibly utilizing each channel. At this stage, definite identification is not possible solely using spectral data. However, this comparison process has vital importance for the signal identification procedure, because the narrow-band spectrum sensing method that will be employed to identify the signal will be selected based on the information provided by SCB. The bandpass filter is applied over time domain data to isolate the signal of interest and estimated bandwidth and center frequency information provided by the wideband sensing stage are used as filter parameters. After the filtering operation, if the SCB information implies possibility of multiple bursts in the filtered data, optional burst detection stage can be utilized for further separation in time domain. Therefore, selection of spectrum sensing method to identify the signal is conducted after the bandpass filter or optional burst detection stage.

One of the ways to classify the narrow-band spectrum sensing methods is based on the signal parameters that are involved in the sensing process and in this context sensing methods are categorized as either coherent or non-coherent techniques \cite{Cabric987}. For instance, matched filtering and cyclostationary feature detection are coherent methods with better detection probability than non-coherent energy detection. However, coherent detectors require a priori information peculiar to the sensed signals: Matched filter provides optimal detection by maximizing signal to noise ratio (SNR) but requires demodulation parameters. Cyclostationary feature detection can detect random signals depending on their cyclic features even if the signal is in the background of noise but it requires information about the cyclic characteristics. But more importantly, while non-coherent methods such as energy detection is applied by setting a threshold for solely detecting the existence of the signal, coherent sensing methods can lead to signal identification. Table~\ref{tab:c1tab1} provides an extensive set of sensing methods classified based on the ability of identification, corresponding hyperspace dimension, and sensing parameters.

\begin{table*}[htb]
\caption{Identification capabilities of spectrum sensing techniques}
\label{tab:c1tab1}
\centering
\begin{tabular}{ c || c | c | c }
\hline
\hline
Sensing Method&Dimension&Sensing Parameter&Result\\
\hline
\hline
Energy Detection&time/frequency&signal energy&Detection Only\\    
\hline
\multirow{3}{*}{Matched Filter}&&time domain signal structure&Detection \\
&time&and characteristics& and\\
&&\textit{e.g.}, pulse shape, package format,& Identification\\
&&guard time, burst duration& \\
\hline
Cyclostationary&&chip rate,data rate,CP size,&Detection\\
Feature&frequency/code&symbol duration, modulation type,& and \\ 
Detection&&carrier spacing and number&Identification\\
\hline
Statistical Tests&time&signal distribution&Detection Only\\
\hline
Entropy Based&frequency&signal entropy&Detection Only\\
\hline
Eigenvalue &time/&signal eigenvalues&Detection Only\\
Based &angle&direction of arrival&\\
\hline
&&cyclic prefix, midamble &Detection\\
Autocorrelation&time&preamble, PN sequence,&and\\
&&and others&Identification\\
\hline
Template&frequency&frequency domain&Detection and\\
Matching&&filter characteristics&Identification\\
\hline
Multitaper Based&frequency&signal energy&Detection Only\\
\hline
Wavelet&frequency&signal energy&Detection Only\\
\hline
Multiband&&&\\
Joint&frequency&signal energy&Detection Only\\
Detection&&&\\
\hline
\hline
\end{tabular}
\end{table*} 

The initial classification information provided by the SCB leads to the possible sensing parameters for potentially occupant signals and based on these information, spectrum sensing method selection block (SSMSB) finds the corresponding sensing method from Table~\ref{tab:c1tab1} and executes identification procedure. In a given scenario, if the SCB indicates the probability of DSSS signals in a given channel, after filtering, SSMSB can employ cyclostationary feature detection from the Table~\ref{tab:c1tab1}. Thus, peak detection block will search for the cyclic frequencies indicated for the peculiar signal to make a decision in the first place but in case of no identification, the range of the cyclic frequency search can be extended. On the other hand, if only the channel occupancy information is important, SSMSB can operate with non-coherent sensing methods. The statistical tests listed in Table~\ref{tab:c1tab1} include but not limited to Anderson-Darling test, student's t-test, Kolmogorov-Smirnov test, and statistical covariance based tests. The last two items in Table~\ref{tab:c1tab1} are also wideband sensing techniques which provide sensing decision directly in contrary to the sweep tune and filter bank based wideband sensing techniques which require another sensing method at the final stage. Modulation classification is kept out of scope of this study because same modulation type and order can be employed by different wireless signals in the spectrum. Thus, identification in the modulation dimension introduces its own challenges. 

Wideband sensing methods that are available in the literature detect the channels which have activity over the spectrum. However, their final binary hypothesis testing process which is based on null and alternative hypotheses does not provide any further information on spectral parameters such as center frequencies and bandwidths of the signals. Moreover, the burst and peak detection stages of the proposed signal identification system model also implies further detection processes. For instance, each coherent narrow-band sensing technique can lead to maximization of a certain parameter to identify a peculiar type of signal and separate maximum likelihood detectors can be designed for each parameter. Such an approach would introduce high level of complexity and would be an inflexible solution. Consequently, there is a need for a comprehensive approach to the wideband sensing and detection requirements of the signal identification process. We address these issues by introducing a modular, reusable noise floor and spectral parameter estimation method.

\section{Noise Floor and Spectral Parameter Estimation Method}
Based on the wideband, multi-channel sensing, spectral parameter estimation, and peak detection requirements of the signal identification system model, we propose a modular noise floor and spectral parameter estimation method (NFSPEM). The proposed NFSPEM first estimates the noise floor and distinguishes the rest of the ``information bearing'' samples from the samples that are marked as noise. Then, the next module estimates the signal parameters such as bandwidth and center frequency. The proposed NFSPEM first achieves sensing of the signals over a given frequency spectrum without the requirement of any a priori information. Moreover in contrary to the other wideband sensing techniques, adopted noise floor estimation approach also leads to the calculation of the spectral parameters such as bandwidth and center frequency which will be utilized at the following stages of the signal identification process. In addition, both burst and peak detection modules can be implemented based on NFSPEM, because a time domain signal or the output of a coherent narrow-band sensing method such as cyclic frequency spectrum exhibit similar characteristics with the frequency spectrum in the context of detection. Hence, the fundamental idea of distinguishing information bearing samples from the noise will work.

In case of initial wideband sensing process, the NFSPEM works on $N$ spectral samples out of FFT. In wideband multi-channel sensing scenarios, it can be assumed that signals of various power levels occupy the spectrum along with noise. The samples with highest power levels will carry the most information about the wireless occupant signals, while the samples that are accumulated at the bottom of the spectrum will be constituted by the noise \cite{Millioz}. Moreover it is known that noise variance is less than signal variance \cite{Kim}, thus it is plausible to divide the samples into two groups: First group of samples starting from the bottom of the spectrum until some level of power should constitute signal base or noise floor and the second group of samples should be evaluated as the information bearing samples for wireless signals. Considering the two groups of samples, it can be stated that the samples that accumulate at the bottom of the frequency spectrum will exhibit a denser distribution when compared to the samples in the second group due to lower variance and consequently lower standard deviation. Therefore starting from the minimum power level, detecting a significant increase in the standard deviation will mark the end of noise floor for the given frequency bands. Such a sensing mechanism can be implemented by adopting the quantization level approach from the analog signal digitization process: The frequency spectrum will be partitioned into horizontal segments and all the samples will be grouped based on their corresponding segments. Therefore, if the power spectral representation of the received signal is defined as $Y(n)$ where $\mathnormal{n} = 1,\cdots,\mathnormal{N}$, the spectrum can be divided into horizontal segments of

\begin{equation}\label{eq:thresh1}
L  = \frac{\max(Y) - \min(Y)}{k \sigma_\mathnormal{Y(N)}},
\end{equation}

\noindent where $\sigma_\mathnormal{Y(N)}$ is the standard deviation of whole $N$-point FFT samples of the frequency spectrum and $k$ is the standard deviation coefficient and can be selected as $0<k\leq 1$ according to adaptive or dynamic threshold methodology which is employed for data binarization \cite{Sezgin}. It should be noted that, the segment length adaptively changes with the standard deviation of the recorded spectrum. If the standard deviation is high, there will be more fluctuations in the frequency spectrum, then, the length of the segments $L$, becomes wider to catch the activity in the sensed frequency bands. If the standard deviation is small, then, spectrum will be relatively flat, segment length becomes narrower, detection and parameter estimation in the next stage become more precise.

The second issue that should be addressed is the selection of quantization levels which will represent the noise. In this case, there is a need to detect the point where a significant change occurs in the standard deviation and consequently in the number of samples in one of the quantization levels due to transition from the first group of samples to the second. When the quantization levels are listed from bottom to top, in case of a significant reduction in the number of samples in the quantization level when compared to the previous level due to high standard deviation of information bearing samples, it can be assumed that the noise floor is limited with the previous level and the samples belong to the level where the change occurs and the samples of other higher quantization levels belong to occupant signals. To this end, when the change detection methods are investigated, cumulative sum (CUSUM) change detection method which is already applied to spectrum sensing in the context of quickest detection \cite{Husheng} becomes a prominent alternative because, in contrary to the Bayes change detection, it does not require any information about the data distribution and it is not time sensitive. If $m_1,m_2,\ldots,m_L$ are the number of samples at each quantization level, the average number of samples per level can be defined as $\bar{m}$ and the cumulative sum starting with $S_0 = 0$ can be given by

\begin{equation}\label{eq:csm2}
S_i = S_{i-1} + (m_i - \bar{m}),
\end{equation}

\noindent where $i = 2,\ldots,L$. After each CUSUM value is calculated, the change point where the quantization levels above the noise floor starts is given by

\begin{equation}\label{eq:csm3}
 S_{max} = \arg\!\max_{i = 2,\ldots,L}S_i.
\end{equation}

The confidence levels of the change point detection method has vital importance from the aspect of sensing performance. Therefore, bootstrap analysis is implemented to quantify the confidence levels of the CUSUM method for wideband sensing. Channel occupancy rate has important influence on the confidence levels. Therefore, the approximate confidence levels for changing SNR and wireless spectrum occupancy rates are given in Table~\ref{tab:411}. The simulation results indicate that as the occupancy rate becomes lower, the confidence levels get slightly better. Actually, when the occupancy rate reaches $90\%$, the system model converges to the single channel systems in which the proposed method provides the same performance with that of energy detection. The Table~\ref{tab:411} indicates that as the SNR $> 10$, a certain level of confidence is achieved for all occupancy rates. When $0\%$ occupancy rate boundary is considered, the bootstrap analysis provides $52.671\%$ of confidence under the condition that all samples are taken from the additive white Gaussian noise.

\begin{table}[htb]
\caption{NFSPEM confidence levels: channel occupancy rate v.s. SNR}
\label{tab:411}
\centering
\begin{tabular}{c || c | c | c | c }
& \multicolumn{4}{c}{ Channel Occupancy Rate}\\
\cline{1-5}
SNR& 25\% & 60\%  & 75\%  & 90\% \\
\hline
\hline
-4 & 36.7480\% &  25.3440\%  & 22.5340\%  & 18.9990\%\\
\hline
-2 & 74.5660\% &  43.6100\%  & 24.5370\%  & 20.4360\%\\
\hline
0 & 94.0610\% &  69.2080\%  & 47.7670\%  & 26.2010\%\\
\hline
2 & 98.7510\% &  92.8520\%  & 73.0890\%  & 40.6680\%\\ 
\hline
4 & 99.6580\% & 98.7490\% & 93.5080\% & 65.5230\%\\
\hline
6 & 99.9070\% & 99.6430\% & 97.9930\% & 86.0050\%\\
\hline
8 & 99.9500\% & 99.8200\% & 99.1730\% & 92.7060\%\\
\hline
10 & 99.9670\% & 99.8300\% & 99.3970\% & 92.9740\%\\ 
\hline
12 & 99.9699\% &  99.9620\%  & 99.7990\%  & 95.9640\%\\
\hline
14 & 99.9700\% &  99.9650\%  & 99.80\% & 96.9830\%\\ 
\hline
16 & 99.9730\% & 99.9680\% & 99.8680\% & 97.1910\%\\
\hline
18 & 99.9740\% & 99.9690\% & 99.8822\% & 97.7850\%\\
\hline
20 & 99.760\% & 99.9690\% & 99.8990\% & 98.4310\%\\
\end{tabular}
\end{table}

\begin{figure*}[t]
\centering
\subfigure[The two sub-blocks of noise floor and spectral parameter estimation method.]{
\includegraphics[width=.4\textwidth]{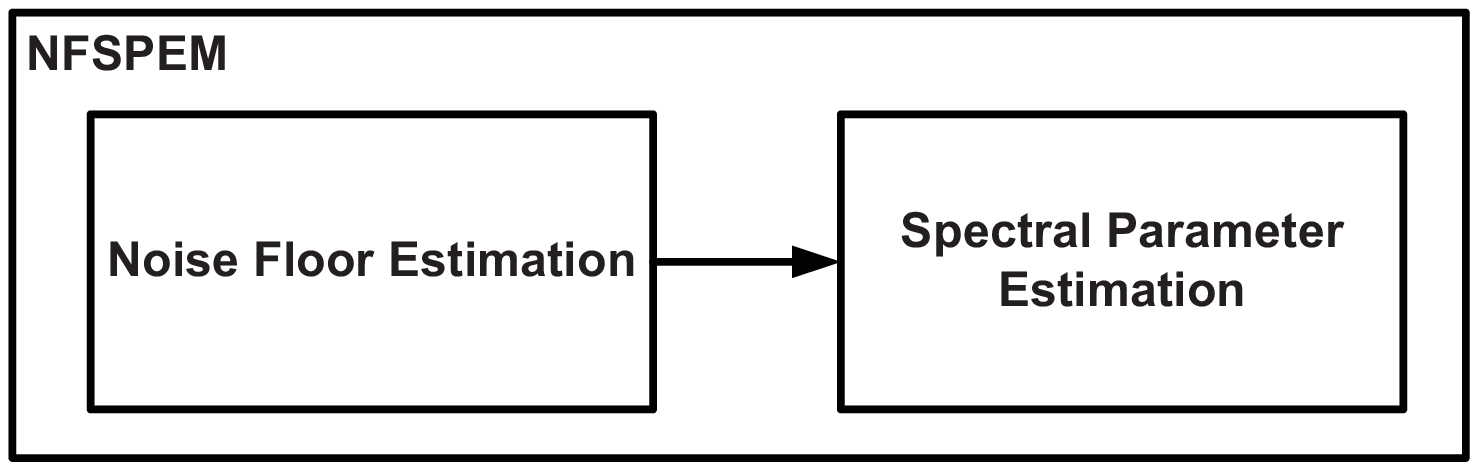}
\label{fig25}
}\\
\subfigure[Wideband sensing at PCS downlink band with NFSPEM: the noise samples which are detected by the first 
sub-block is marked with red. Greed diamonds indicate estimated center frequencies by the second sub-block.]{
\includegraphics[width=.5\textwidth]{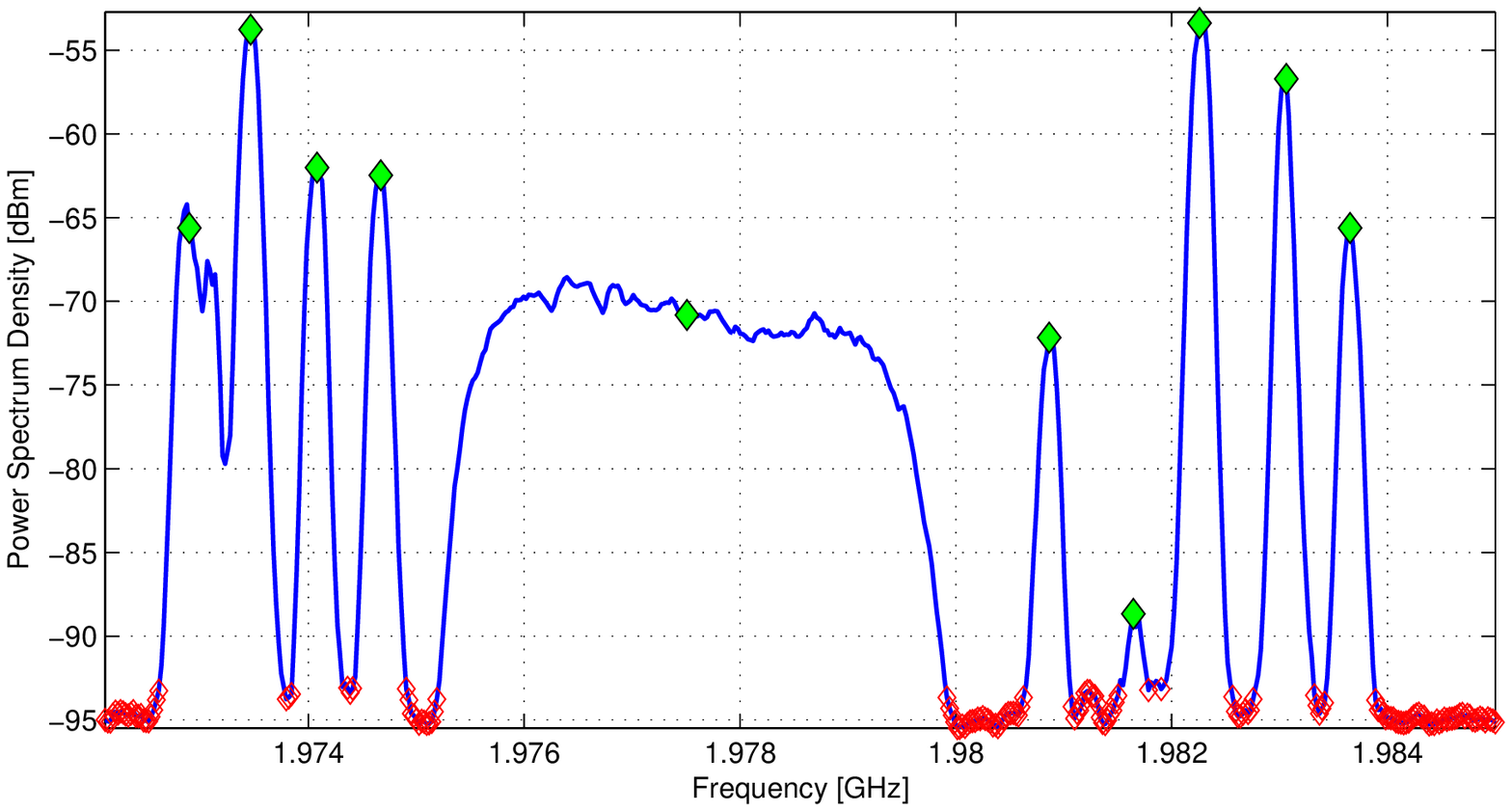}
\label{fig19}
}\\
\subfigure[Burst detection with NFSPEM: noise floor is marked with red and second sub-block estimates the burst times and durations in an ISM band recording snapshot.]{
\includegraphics[width=.5\textwidth]{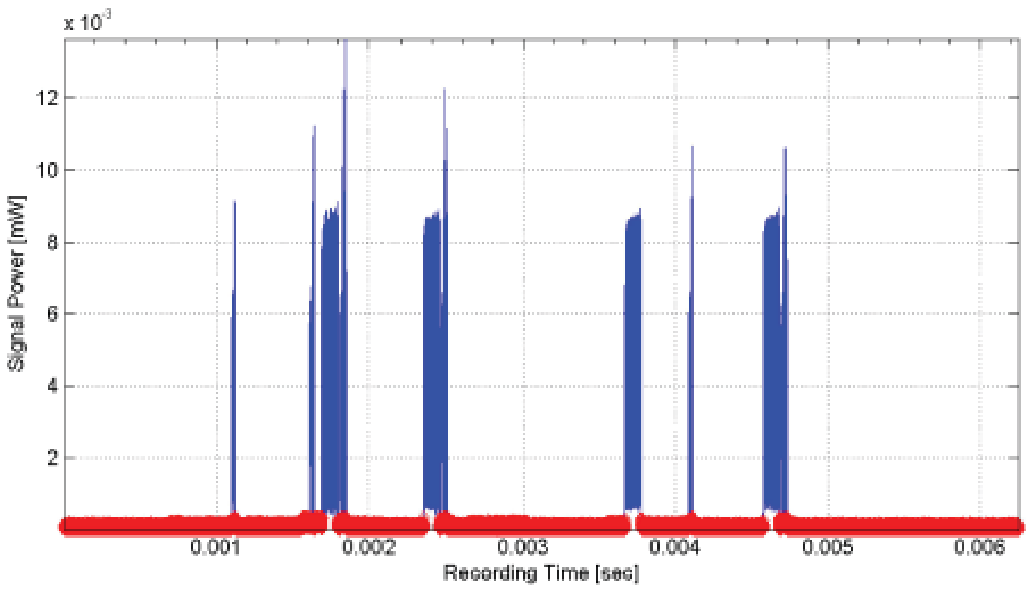}
\label{fig20}
}\\
\subfigure[Peak detection with NFSPEM: cdma2000 signal identification and carrier number estimation. The NFSPEM provides the dominant peaks at the cyclic spectrum.]{
\includegraphics[width=.5\textwidth]{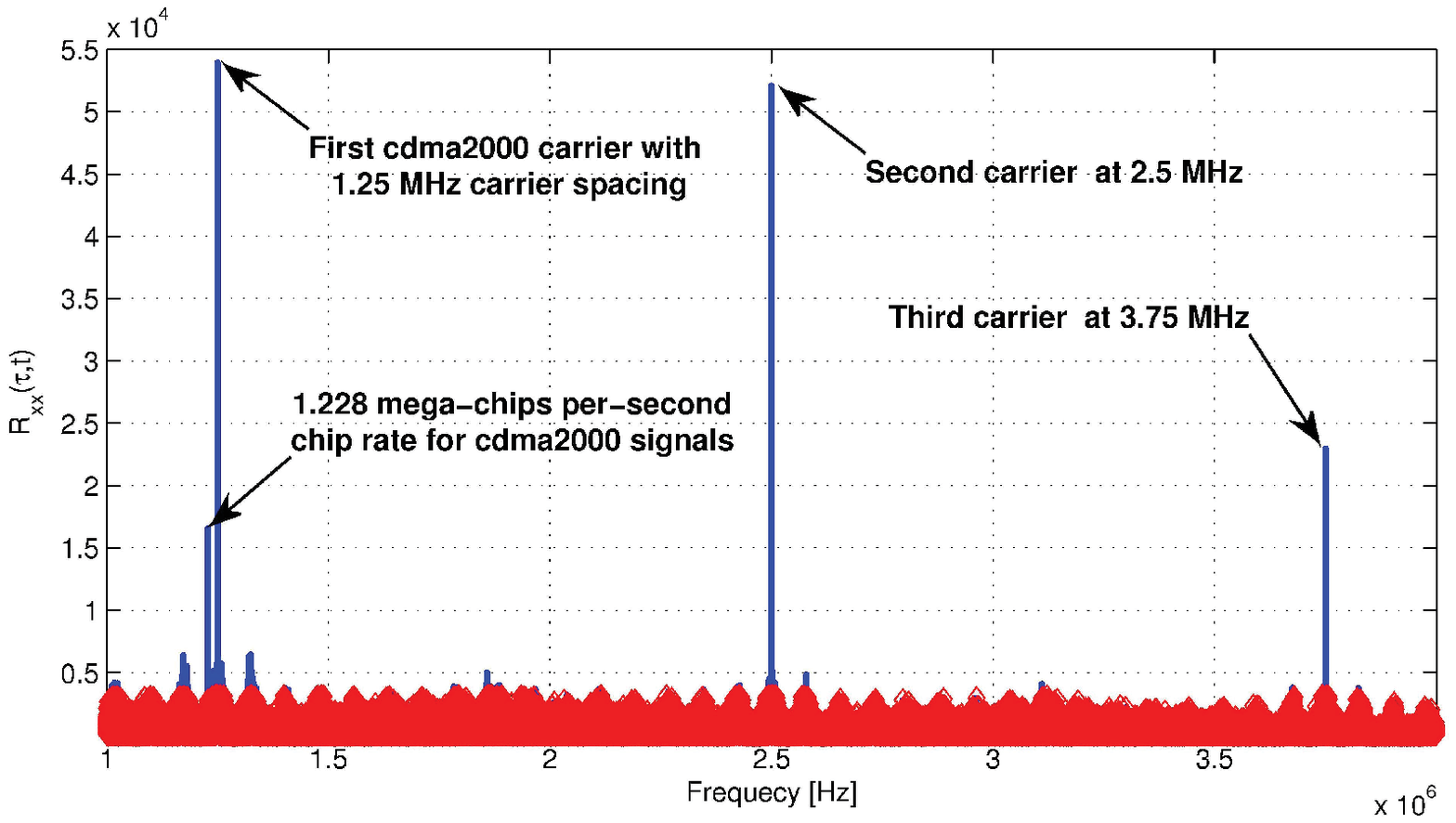}
\label{fig32}}
\caption{The modular NFSPEM can be utilized for wideband sensing, burst and peak detection requirements of signal identification system model. Reusing NFSPEM leads to significant complexity reduction.}
\label{c1fig4}
\end{figure*}

Beside wideband sensing, the NFSPEM can be utilized for time domain burst and other peak detection requirements of the signal identification. Reusing NFSPEM for such functionalities leads to significant reduction at the implementation requirements and complexity of the system model. The time domain recording or the output of a signal processing application such as autocorrelation or cyclic spectrum can be given as inputs to the NFSPEM. After the information bearing samples are distinguished, the second parameter estimation sub-block of the method provides a set of parameters such as burst duration and peak location. These information is utilized either by the SSMSB to chose the narrowband sensing method or by the decision mechanism which will lead to identification directly. The sub-blocks of NFSPEM are shown in Fig.~\ref{c1fig4} along with example measurement outputs for wideband sensing, burst, and peak detection. The peak detection case corresponds to identification and parameter estimation of a 3-carrier cdma2000 signal. While the first peak identifies the unique chip rate of IS-95/cdma2000 signals \textit{i.e.,} $1.228$ mega-chips per-second, the following peaks provide the total number of carriers and their spacings.

\section{Modular Flexibility of System's Blocks}
On the way to realizing fully adaptive, aware, and intelligent wireless communications systems, signal identification should help to achieve the itemized goals of ASHA in the first place. Therefore, starting from the RF front-end, signal processing, detection, estimation and decision making blocks of signal identification system model should be considered as exchangeable and updatable components. Whenever new spectrum sensing methods or identification algorithms are developed, the spectrum sensing block which comprise available sensing methods, the SSMSB, and consequently Table~\ref{tab:c1tab1} should be updated. Based on the requirements of newly introduced sensing techniques, a peculiar set of decision mechanisms can emerge, thus the decision mechanism should also be modified accordingly. Moreover, spectral parameter estimation, burst and peak detection requirements of the proposed system model can be satisfied with robust, comprehensive, and computationally sound detection and estimation methods that can be proposed later. On the other hand, different noise floor estimation methods can be employed for the specific needs of different scenarios. For instance, if the noise estimation precision is highly important, in return for additional computational complexity, a possible increase at the sensing duration, and a newly designed decision mechanism, spectral kurtosis based noise variance estimation method proposed in \cite{Millioz} leads to approximately $3\%$ improvement at $25\%$ channel occupancy rate and $5\%$ improvement at $75\%$ occupancy for SNR $< 10$ dB when compared to the proposed noise floor estimation sub-block. Therefore current solutions can be exchanged with the new ones whenever necessary, as a part of continuous improvement efforts to reach the objectives of ASHA.

\section{Application Areas}
One of the potential application areas for the proposed system is the identification and location estimation of illegal emitters \cite{SM.1600-1}: Signal identification systems can provide information such as the signal type and direction of arrival to direct the regulatory officials and field engineers to the source of abusive usage. Beside that there can be some unintentional emissions in the communications environment caused by uncalibrated or broken access point or base station circuitry such as power amplifiers and oscillators. It can be problematic to find the source of such transmissions due to the number and distribution of the transceivers over the operation area. However, signal identification algorithms can directly find emitting sources or quickly reduce the number of candidate transmitters to a few. 

The signal identification procedure in Fig.~\ref{c1fig3} can be utilized by numerous governmental, commercial, and military applications. The potential applications are frequency management, security and surveillance, interference management, geolocation of target emitters. Therefore, signal identification can be beneficial to frequency regulation agencies, public safety agencies, cellular operators, broadcasters, transportation agencies for navigation and communication, and law enforcement. For instance, TV white space systems aim to provide wireless wideband access through unused radio spectrum at the TV channels. If the vacant channels would be decided based on only the TV white space spectrum databases, in border regions, broadcasts from the neighboring county or city could be assumed insignificant source of interference while sometimes they would be. In such scenarios wireless access QoS and broadcasting quality would degrade. Therefore TV white space systems can benefit from the signal identification system for precise TV white space access planning. Moreover, signal identification can lead to information regarding to the co-channel, adjacent channel, narrowband, and wideband interference sources. IEEE 802.16 Broadband Wireless Access Working Group working on the interference detection and measurement indicates that ``based on the above conditions, an accurate detection and measurement of the intranet interference requires specific interference patterns to be evaluated across a given cluster of cells subject to the interference detection and measurement''. Even though this document is standard specific, it defines the ``interference pattern" concept which binds the modeling and prediction efforts to the interference detection problem. Thus, interference analysis methods can benefit from the outputs of the signal identification procedures as well as spectrum modeling efforts.

One very important field that signal identification methodology can contribute is the public safety communications. Emergency call services such as enhanced 911 are designed to provide improved service including instant delivery of the victims location information to the local Public Safety Answering Point (PSAP). Taking the requirements of the 911 services into account, wireless service providers took incentive to employ precise location estimation techniques based on their network capacities and structures. Even though these systems are very helpful for limited number of calls and specific events, it is not possible to maintain such services under extreme cases affecting an important section of the population living in an area. The disasters such as Pakistan and Australia vast area floods, California wild forest fires, and earthquakes in China and Japan showed that the first hours and days are very crucial in saving the lives of the victims. In this important period of time, the victims who are scattered around the disaster area would be looking for a way to communicate to get help. However, the wireless network infrastructure can be damaged during such extreme situations. Beside that, if the victims are calling to reach the 911 services at the same time, congestion due to the capacity limit of either network or PSAP can result latency or it may not even be possible to provide most of the victim location information to the public safety officers in a timely manner. In such cases the signals transmitted by the devices of the victims can be detected, tracked and the direction of the signals can be estimated by the deployed ad-hoc signal identification hardware located around the disaster areas. Moreover, the signal detection, location estimation and direction finding algorithms can be developed in a flexible manner. These technologies can be employed by the first responders to detect and locate victims even for cases in which the original core wireless communications network is down. The first responder teams and their equipments can be used to detect the victim transmission which can lead to locate the victim transmitter via direction finding and location estimation algorithms.

\section{Open Issues}
Identification duration and accuracy which replace sensing duration and accuracy in the context of proposed system model are the main parameters to quantify the efficiency of opportunistic access to the spectrum hyperspace. The rapid changing communications medium requires the signal identification to be conducted as quick as possible with a certain level of confidence and in general, the value of both parameters increase with increasing algorithm and signal processing complexity. The proposed signal identification system model provides more information about the hyperspace when compared to the spectrum sensing methodology as it establishes itself over the current state of the art by employing both narrow-band sensing methods and wideband sensing as sub-blocks. Gaining extra knowledge about the communications medium and extending spectrum sensing to the dimensions of the spectrum hyperspace consequently entailed the implementation of additional detection, classification, and sensing algorithms along with signal processing components such as bandpass filtering. Furthermore, sequential execution of identification procedures defined in Fig.~\ref{c1fig3} over the monitored wideband spectrum leads to extension of identification duration. Implementation of parallel filtering architectures such as filter banks along with a multiple antenna RF-front end such as the one introduced in \cite{Cabric987} can lead to identification of opportunities quicker than the sequential process, however, the additional system level complexity that will be introduced should be quantified carefully.
 
AGC block can benefit from the extensive frequency domain occupancy modeling and prediction research which is classified in \cite{Gorcin} to establish balance between sampling rate and dynamic range requirements in an intelligent manner. Another issue is the selection of the wideband spectrum to be monitored. Instead of implementation of multiple antennas, RF front-end can also be designed in a flexible fashion and can be tuned to different chunks of spectrum. Wideband channel selection strategies introduced in \cite{Li} can be helpful in such an architecture. On the other hand, after the initial signal identification tasks are completed successfully, the prosed ASHA methodology can benefit from the extensive research on multi-user detection methods to further determine available communications opportunities at the user level in the focused dimension. The design of waveforms can be conducted based on these information and access to the spectrum hyperspace can be executed in the context of ASHA. Signal identification processes can also benefit from cooperative communication architectures by distributing the work load between the wireless nodes in communication. Collaboration scenarios similar to the spectrum sensing can be discussed for the signal identification as well.

\ifCLASSOPTIONcaptionsoff
  \newpage
\fi

\end{document}